\documentclass[sigconf]{acmart}
\AtBeginDocument{%
  \providecommand\BibTeX{{%
    \normalfont B\kern-0.5em{\scshape i\kern-0.25em b}\kern-0.8em\TeX}}}
\setcopyright{acmcopyright}
\copyrightyear{2024}
\acmYear{2024}
\setcopyright{acmlicensed}
\acmConference[CHI '24]{Proceedings of the CHI Conference on Human Factors in Computing Systems}{May 11--16, 2024}{Honolulu, HI, USA}
\acmBooktitle{Proceedings of the CHI Conference on Human Factors in Computing Systems (CHI '24), May 11--16, 2024, Honolulu, HI, USA}
\acmDOI{10.1145/3613904.3642237}
\acmISBN{979-8-4007-0330-0/24/05}

\usepackage{color}
\usepackage{tabularx}
\usepackage{hyperref}
\usepackage{enumitem}

\definecolor{themeblue}{rgb}{0.14, 0.52, 0.87}

\newcommand{\highlight}[1]{\textcolor{black}{#1}}

\newcommand{\add}[1]{\textcolor{black}{#1}}
\newcommand{\workshop}[1]{{#1}}
\newcommand{\userstudy}[1]{{#1}}
\newcommand{\rr}[1]{\textcolor{black}{#1}}
\newcommand{\rrindex}[1]{}

\begin{document}

\newcommand{\system}{VAID}

\title[VAID: Indexing View Designs in Visual Analytics System]{VAID: Indexing View Designs in Visual Analytics System}

\author{Lu Ying}
\email{yingluu@zju.edu.cn}
\orcid{0000-0002-4206-231X}
\affiliation{
  \institution{Zhejiang University}
  \city{Hangzhou}
  \state{Zhejiang}
  \country{China}
}

\author{Aoyu Wu}
\email{aoyuwu@g.harvard.edu}
\orcid{0000-0001-9187-9265}
\affiliation{
  \institution{Harvard University}
  \city{Cambridge}
  \state{Massachusetts}
  \country{United States}
}

\author{Haotian Li}
\email{haotian.li@connect.ust.hk}
\orcid{0000-0001-9547-3449}
\affiliation{%
  \institution{The Hong Kong University of Science and Technology}
  \city{Hong Kong SAR}
  \country{China}
}

\author{Zikun Deng}
\email{zkdeng@scut.edu.cn}
\orcid{0000-0002-4477-5292}
\affiliation{%
  \institution{South China University of Technology}
  \city{Guangzhou}
  \state{Guangdong}
  \country{China}
}

\author{Ji Lan}
\email{jilan2-c@my.cityu.edu.hk}
\orcid{0000-0002-8658-8620}
\affiliation{
  \institution{AIFT}
  \city{Hong Kong}
  \country{China}
}

\author{Jiang Wu}
\email{wujiang5521@gmail.com}
\orcid{0009-0001-8831-1473}
\affiliation{
  \institution{Zhejiang University}
  \city{Hangzhou}
  \state{Zhejiang}
  \country{China}
}

\author{Yong Wang}
\email{yongwang@smu.edu.sg}
\orcid{0000-0002-0092-0793}
\affiliation{
  \institution{Singapore Management University}
  \country{Singapore}
}

\author{Huamin Qu}
\email{huamin@cse.ust.hk}
\orcid{0000-0002-3344-9694}
\affiliation{%
  \institution{The Hong Kong University of Science and Technology}
  \city{Hong Kong SAR}
  \country{China}
}

\author{Dazhen Deng}
\authornote{Dazhen Deng is the corresponding author.}
\email{dengdazhen@zju.edu.cn}
\orcid{0000-0002-9057-8353}
\affiliation{
  \institution{Zhejiang University}
  \city{Hangzhou}
  \state{Zhejiang}
  \country{China}
}

\author{Yingcai Wu}
\email{ycwu@zju.edu.cn}
\orcid{0000-0002-1119-3237}
\affiliation{
  \institution{Zhejiang University}
  \city{Hangzhou}
  \state{Zhejiang}
  \country{China}
}

\renewcommand{\shortauthors}{Trovato and Tobin, et al.}

\begin{abstract}
    Visual analytics (VA) systems have been widely used in various application domains. However, VA systems are complex in design, which imposes a serious problem: although the academic community constantly designs and implements new designs, the designs are difficult to query, understand, and refer to by subsequent designers. To mark a major step forward in tackling this problem, we index VA designs in an expressive and accessible way, transforming the designs into a structured format. We first conducted a workshop study with VA designers to learn user requirements for understanding and retrieving professional designs in VA systems. Thereafter, we came up with an index structure VAID to describe advanced and composited visualization designs with comprehensive labels about their analytical tasks and visual designs. The usefulness of VAID was validated through user studies. Our work opens new perspectives for enhancing the accessibility and reusability of professional visualization designs.
\end{abstract}



\maketitle

\section{Introduction}
Visual analytics (VA) combines data mining and visualization techniques to help users with data exploration in different domains, such as biology~\cite{krueger2019facetto, lekschas2018hipiler}, sports~\cite{stein2018bring, cao2021migviewer}, urban~\cite{deng2023survey, lu2023waterexcva}, and explainable AI~\cite{gou2020vatld, ono2021pipeline}.
Researchers in VA have developed advanced VA systems with highly customized visualization designs for obtaining insight into data~\cite{sacha2014knowledge}.
Designing effective VA systems is highly challenging and demanding, requiring close collaboration between experienced visualization practitioners and domain experts.

Views are basic building blocks of VA systems.
To create an effective VA system, it is critical to design views by mapping the data and tasks derived from domain problems to visual designs~\cite{munzner2009nested}.
Recent advance in visualization has attempted to automate such a mapping process~\cite{srinivasan2018augmenting, deng2022dashbot, chen2020composition}.
However, these studies recommend basic statistical charts (e.g., bar and line charts) for low-level analytical tasks such as finding distributions.
They can hardly support designing views in VA systems that deal with complex datasets and tasks~\cite{keim2008visual}.
\rrindex{O2}\add{The existing VA design pipeline heavily relies on researchers' experience, requiring surveying related studies and summarizing design requirements for new scenarios.
Passing examples can offer valuable inspiration due to the multitude of design styles in existence~\cite{lee2010designing, herring2009getting, bigelow2014reflections}. 
To better support the designers and researchers of visual analytics, inspired by research in creativity support~\cite{shneiderman2002creativity}, we believe it is important to facilitate the ideation process by enabling the exploration of previous VA view design.}
However, without an effective indexing method, currently, these view designs can only be searched through simple keywords, such as domain problems, which cannot fulfill the requirements of VA designers.
Unable to search with more fine-grained requirements, they struggle to draw inspiration from a large number of prior successful designs.

To address the challenge, we aim to propose an indexing approach for view designs in VA systems collectively considering \textbf{tasks}, \textbf{data}, and \textbf{visualizations}~\cite{munzner2009nested}.
However, it is unclear how to define an index structure based on these factors.
For example, from the visualization perspective, a VA design might contain a hybrid use of different visual elements, such as composite visualizations~\cite{javed2012exploring} and glyphs~\cite{borgo2013glyph}.
When representing these complex designs with indexes, preserving all details can lead to difficulties for designers in specifying their searching criteria and understanding their structure and semantic meanings.
On the other hand, if the information is over-abstracted to a high-level description, such as a few keywords, designers may struggle to accurately express their design when searching for required design information from indexes of returned visual designs.
It is important to balance expressiveness and efficiency when designing the index structure. 
Therefore, to understand designers' requirements on the index structure, we conducted a workshop study with 12 VA designers, most of whom have published papers in the IEEE VAST conference as the first author. 
With the study, we validated the necessity of indexing past designs for creating new ones and collected the requirements for constructing such an index.

Based on the user feedback from the workshop study, we formulated an index structure named \system{} for VA designs inspired by Vega-Lite~\cite{satyanarayan2016vega}.
\system{} enables an expressive characterization of visualizations with analytical tasks and visual designs.
To ensure the coverage and comprehensiveness of the index, we iteratively labeled the view designs and refined the keys and values of the index structure.
As a result, we gained 442 view designs from 124 VA systems and formed an informative index structure for them.
To demonstrate the usefulness of \system{}, we conducted a user study using a prototype for the exploration of \system{} with 12 participants.
Specifically, we asked participants to query designs for specific analytical tasks, data types, mark types, etc.
User feedback showed that \system{} could help them query diverse and useful visual designs, thereby aiding their design exploration.
Leveraging the usefulness of VAID in presenting view designs, we proceeded to conduct an in-depth analysis and obtained findings into the patterns of VA view designs. 
Finally, we concluded our research by discussing future directions and limitations.
The contributions of this paper include:
\begin{itemize}
    \item
    requirements for understanding and indexing views in VA derived from a workshop study;
    \item an effective index structure \system{} for VA designs (including analytical tasks and visual designs);
    \item a user study based on an exploration prototype\footnote{\url{https://VIS-VAID.github.io/}} to demonstrate the usefulness of \system{};
    \item an in-depth analysis of existing view designs and research opportunities based on VAID.
\end{itemize}

\section{Related Work}
This paper is related to studies about visualization indexing, visual analytics design studies, and visualization typologies.

\subsection{Visualization Indexing}
Given that visualizations usually have complex structures of visual components, numerous studies investigate the indexing and searching of visualizations.
One way for indexing is by assigning tags to the visualizations.
Many visualization datasets collect visualizations and categorize them by types, such as MASSVIS~\cite{borkin2013makes}, VizNet~\cite{hu2019viznet}, VIS30K~\cite{chen2021vis30k}, VisImages~\cite{deng2023visimages}, and Many Eyes~\cite{Viegas2007manyeyes}.
Tagging visualizations is useful for machine learning model training, but the tags have limitations when it comes to analyzing visualization configurations. 
Important configurations like visual encodings, compositions, and associated tasks are crucial for comprehending the designs of visual analytics, and these aspects go beyond the scope of traditional tags.

Computational methods have been used to extract and index visualizations.
For example, when retrieving SVG charts, to ensure both the similarity of visual structures and data distributions, Li et al.~\cite{li2022structure} proposed a method based on graph neural networks for feature modeling.
Hoque et al.~\cite{hoque2019searching} collected visualizations that are implemented by D3.js and parsed the hierarchical structures of the visualizations.
However, parsing and analyzing bitmap charts is a more challenging task compared to SVG charts.
A series of methods adopt computer vision methods to reverse-engineering visualizations~\cite{savva2011revision, poco2017reverse, ying2023reviving, zhou2023intelligent} or extracting numerical representations for charts indexing~\cite{ye22022visatlas}.
Though effective, these methods might not be applicable to the charts in visualization publications, which usually have complex layouts and composite designs.
In this work, we focus on analyzing visualizations in the context of visual analytics, which poses higher requirements for data labeling. Specifically, it not only requires labeling the meta information like chart positions but also the information related to visualization literacy (e.g., visual encodings and tasks).
Our efforts form a valuable index structure of visualization designs from state-of-the-art VA systems.

\subsection{Visualization Design in Visual Analytics}
Since the analysis problems and data structures are getting more complex in recent years, visual analytics (VA) systems are equipped with more features to fulfill the analytical requirements.
Therefore, researchers reflected on the scope and challenges of VA~\cite{Keim2008, kui2022surveya} and proposed a series of conceptual models.
For example, Sacha et al.~\cite{sacha2014knowledge} have proposed a knowledge generation model to characterize VA systems and their use in sensemaking.
According to the model, VA systems should be well integrated into the human knowledge generation loop from hypothesis and action to derive the findings and insights.
Moreover, they think that the VA system is composed of three components, namely, data, visualization, and algorithms, involving the pipeline of information visualization and the process of knowledge discovery and data mining. 

To design visualizations that are compatible with the knowledge generation pipeline~\cite{sacha2014knowledge}, Munzner~\cite{munzner2009nested} has proposed a nested model for visualization design and evaluation.
The nested model consists of four stages: 1) domain problem and data characterization, 2) operation and data type abstraction, 3) visualization design, and 4) algorithm design.
The first two stages are considered different levels of abstraction of the data and tasks.
With the abstracted data and tasks, the design choices of visualizations can be further derived based on the theories in information visualization, such as expressiveness and effectiveness criteria~\cite{mackinlay1987automatic} and the rules of visual mapping~\cite{card1999readings,munzner2014visualization}.
The nested model provides prescriptive guidance for visualization experts in constructing VA systems.
Inspired by the model, we construct an index structure of visual analytics describing visualizations from their analytical tasks and visual designs.
Compared to conceptual models, the structure we present is a unique contribution to the community for data-driven analysis and design inspiration to promote research on VA systems.

\subsection{Visualization Taxonomy and Grammars}
The classification of visualizations~\cite{harris1999information,chi2000taxonomy, lohse1994classification, engelhardt2018framework, meirelles2013design} has been studied for a long time.
For example, Borkin et al.~\cite{borkin2013makes} classified visualizations into 12 categories, such as \textit{Area}, \textit{Bar} and \textit{Circle}, each comprising multiple sub-types.
However, the designs of visualizations for visual analytics usually have novel layouts and complex compositions.
Chen et al.~\cite{chen2020composition} have attempted to map each view in VA systems to a specific visualization category in Borkin's taxonomy.
However, they discovered that a view might be ambiguous to a specific category because many designs contain various visual components of multiple categories.
They reflected on the categorization and proposed to follow Javed et al.'s theory of composite visualization~\cite{javed2012exploring} to characterize the visual designs in further research.
Based on this reflection, we regard the visualizations in VA systems as composite visualizations and characterize the relations between the components with a hierarchical specification.

Grammars of graphics~\cite{wilkinson2012grammar} are fundamental in visualization systems, indicating the visual mappings from data to visual channels and layouts.
Mackinlay~\cite{mackinlay1987automatic} formulated visualizations as a graphical presentation problem and adopted relation tuples to specify the data features and visual encodings.
Heer et al.~\cite{heer2010declarative} proposed using declarative languages to describe and specify the visualizations, which is intuitive for the programmers.
After that, Bostock et al.~\cite{bostock2011d3} delivered D3, a programming grammar to operate on the graphical elements of document object model pages.
To further reduce the burden of visualization specification, Satyanarayan et al.~\cite{satyanarayan2016vega} proposed Vega-Lite, a JSON-based declarative programming language, by which users can render a visualization with even several lines of JSON text.
After that, specifying visualization using JSON files becomes widely used, and similar languages are evolving, such as ECharts~\cite{li2018echarts}.
In this paper, we refer to Vega-Lite and extend its style to support the indexing of view designs in VA systems.

\section{Preliminary Study}
\label{sec:workshop}
We conducted a workshop study with VA designers to 
1) understand whether reviewing state-of-the-art visual designs can help visualization designers in design inspiration and 
2) obtain the requirements for understanding and indexing VA view designs.

\subsection{Data Preparation.}
\label{sec: data preparation}
Before the study, we first prepared the state-of-the-art VA designs and derived an initial index design based on tasks, data, and visualizations~\cite{sacha2014knowledge}.

\textbf{Collecting Figures.}
We first collected the figures of VA designs based on VisImages~\cite{deng2023visimages}, which consists of bitmap images collected from IEEE InfoVis and VAST, the top venues for visualization and visual analytics. 
We chose the papers in the IEEE VAST from 2016 to 2020, the primary venue for VA research \rr{(253)}.
Then we identified the papers that propose visual analytics systems, whose paper types are commonly referred to as applications or design studies \rr{(124)}.
For each paper, we selected one figure containing the complete system interface, which was usually the teaser.

\textbf{Separating Individual Visualization Views.}
We further separated the area of different views in the system figures. 
In most cases, a view is assigned a specific name for identification. However, a view sometimes consists of several independent sub-views. If the data among sub-views are not directly related (such as sharing the axes or connected with visual links), we decompose the view into sub-views for different visualizations.
Each sub-view is regarded to be an individual visualization and is the basic analysis unit throughout the paper.
After the annotation, we obtained an image collection of 442 views from 124 VA systems.
\textbf{For simplicity, the term ``view design'' in the rest of our paper refers to the design of a view in a VA system in default.}

\textbf{Annotating Task/Data/Type.}
Derived from VisImages, the views include information about the chart types and their positions but do not provide labels for the views (including sub-views) within VA systems.
We first characterize the view designs by tuples of their task types, data types, and visualization types.
\begin{itemize}
    \item For the \textbf{task type}, we used a task taxonomy of data analysis~\cite{amar2005low} for annotation.
    The taxonomy contains ten types including \textit{retrieve value}, \textit{derive value}, \textit{filter}, \textit{find extremum}, \textit{sort}, \textit{determine range}, \textit{characterize distribution}, \textit{find anomalies}, \textit{cluster}, \textit{correlate}, and \textit{compare}.
    To avoid bias during the annotation, we identified a task only when the original authors had mentioned it explicitly. 
    After the annotation, 98.87\% (437/442) visualizations contain at least one task type.
    \item For the \textbf{data type}, we represent the encoded data by its types in visual encoding channels.
    The data types include quantitative (Q), temporal (T), ordinal (O), nominal (N), and graph-related (G) data~\cite{satyanarayan2016vega}.
    For a view design, we summarize the counts of each data type, such as ``$Q\times1, N\times2$.''
    \item We further labeled the \textbf{visualization types} of each visualization.
    For most views, we adhere to the labels used in VisImages, as they were originally assigned based on the taxonomy outlined by Borkin et al.~\cite{borkin2013makes}.
    For a composite visualization, we characterize its type by decomposing it into several visual components.
    For example, a scatterplot matrix can be considered as nesting scatterplots into a matrix, which is represented as a tuple: ``$(scatterplot, matrix)$.''
\end{itemize}

Based on the annotating result, we created a prototype named \system{}-Alpha. 
This interface includes the collected figures, associated tasks, data, and their respective types as searchable indexes. 
Additionally, it features a search engine for direct access.

\subsection{Study Setup}
In the workshop study, we asked participants to imitate the process of designing visualization prototypes, with a specific focus on the task of creating multiple views to accomplish VA tasks. 
We followed the think-aloud protocol and gathered qualitative feedback from participants.

\textbf{Problem.} 
We selected mini-Challenge 2 from IEEE VAST Challenge 2021, a classic problem in the visual analytics field, due to the need for tasks and datasets with an appropriate level of complexity.
Specially, a company, GAStech, hopes to investigate employees' potential private use of the company cars. 
To facilitate analysis, the GPS data of each car, records of car assignments, and records of credit card and loyalty card purchases are provided. 
In our study, the participants are required to achieve three tasks by designing visualization prototypes. 
The first task (\textbf{T1}) is designing visualizations with only credit card and loyalty card data to identify the popular location and purchase time and potentially discover some anomalies (e.g., weird purchase time and frequent changing purchase location). 
The second task (\textbf{T2}) is using car assignment data and GPS data to help determine the owner of each card and trying to find some anomalies (e.g., the card owner and purchase activity are not in the same place). 
The third task (\textbf{T3}) is to reveal the potential unofficial relationships between the employees.

\textbf{Participants.} 
We recruited 12 VA designers \rrindex{S3.1.3}\add{from social media
and our networks}. The participants are postgraduate students (6 females and 6 males) with a research interest in visual analytics for various domains, such as digital humanities, sports analysis, medicine, and urban planning. 
Ten of them have published papers in IEEE VIS as the first author.
We asked participants to report on their experience in designing visualizations for data analysis.
Based on the pre-study interview, 6 participants (\workshop{P1, P3, P4, P8, P10, P11}) are Ph.D. students who have more than three years of research for visual analytics, while 3 (\workshop{P2, P9, P12}) have less than one year and the remaining 3 participants have less than two years. Specifically, 3 participants (\workshop{P1, P3, P12}) own bachelor’s degrees in design.

\textbf{Procedure.} Each trial of the workshop study was conducted one-on-one \rrindex{S3.1.3}\add{via online meetings}. 
A trial lasted about 60 minutes. Before a trial, we first asked participants for their agreement to collect their design process, comments, and results for research use. After that, a study session started with a 15-minute tutorial introducing the prototype \system{}-Alpha's indexes (i.e., how we define the data representations, task types, and visualization types) and several examples illustrating how to use the interface. 
During the study, participants were asked to explain how they understood the problems and tasks and what and why visualizations they wanted to design. The participants needed to sketch the prototype visualizations on paper and illustrate how to use the designs to accomplish the tasks. 
\rrindex{S2.1.1}\rr{The session ended with post-study interviews where we gathered qualitative feedback from participants through a series of questions.}

\subsection{Results}
\rrindex{S1.1}\add{
The overall reactions from the participants were positive. Here we summarize the participant feedback.}
\subsubsection{Usefulness Analysis}
From participant feedback, we found that VAID-Alpha is useful for inspiring new design ideas and enhancing users' original ideas. 
To understand the effect on the design, we further analyzed the design process of the users including system logs, audio recordings, and notes. 
In total, we obtained 36 visualization designs from 12 participants, with 16 (44.4\%) inspired from scratch and 14 (38.9\%) designs enhanced from an original idea. 
The numbers also conform to the participant feedback, demonstrating the usefulness of our system.

\add{\textbf{The system facilitates ``warm-start''.}
We discovered that more participants (\workshop{P2, P5, P7, P8, P10, P11, P12}) are inspired when working on T1 compared to T2 and T3. 
This might be due to the ``cold start'' of the visualization design process,
that is, 
contributing a prototype from scratch requires inspiration and therefore might be difficult at the beginning.
The comments from \workshop{P5}, a junior PhD student, evidenced this inference: ``\textit{at the beginning, I have no idea about the design. Therefore, I preferred to explore and find some inspiration from the recommendations}.'' 
After finishing T1, her design for T2 followed the previous idea with some enhancement provided by the recommendations. 
\workshop{P2, P7, P12}, as junior researchers, also shared a similar design process. 
Besides, 
senior PhD students (\workshop{P8, P10, P11}) also tend to gain inspiration at T1. 
\workshop{P10} commented that he was inspired by a design similar to ``storyline'' after searching with data types and came out with the original design.}

\add{\textbf{The system should help users understand the design.}
Users also expressed concerns about comprehension.
Both junior (\workshop{P2, P3, P8}) and senior (\workshop{P10, P11}) PhD students encountered problems understanding the view designs. 
\workshop{P2} noted, ``\textit{I need to understand the visual encoding for different designs,}'' appreciating textual explanations about data structure and task types but still finding some complex designs challenging to comprehend.
\workshop{P11} also thought that the contextual information provided by the caption was insufficient.}

\subsubsection{Trade-offs between Data, Task, and Visualization}

\begin{figure}[tb]
    \includegraphics[width=\linewidth]{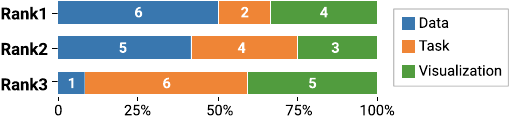}
    \caption{
    The frequency of users' preference rankings for data, tasks, and visualization. For instance, 6 participants ranked ``Data'' as their top choice.
    }
    \label{fig:preference}
\end{figure}

During the study, we also surveyed users' choices for data, tasks, and visualizations for searching visualization designs of interest. 
The results are shown in Fig.~\ref{fig:preference}.

\textbf{``Data'' is the most preferred.} 
We discovered that 6 out of 12 participants ranked the data first, and five participants ranked it second. 
The participants all agreed that data is the most critical factor to consider during visualization design. 
P4, a senior urban planning analyst, commented that ``\textit{data and algorithms are the most critical from the perspective of an expert}''. 
The other two seniors, \workshop{P10} and \workshop{P11}, held a similar opinion. 
\workshop{P5}, a junior, also regarded the data as her priority consideration for visual design, saying that ``\textit{the same data could be represented with different visual representations}''.
However, knowing how many columns with different data types are encoded seems insufficient.
\workshop{P10} commented, ``in real scenarios, data transformation would be performed, and it might be more important to tell how the data are mapped to the visual channels''.

\textbf{``Visualizations'' are more preferred by designers.}
Four participants ranked visualization first, and three of them (\workshop{P1, P3, P12}) are senior designers with bachelor's degrees in design. 
An advantage of searching by visualization types is the consistency between the expectations and outputs.
\workshop{P1} liked searching by visualizations because ``\textit{visualizations are very intuitive for understanding and I can imagine what results will appear}''. 
Among three dimensions, \workshop{P3} highlighted his preference for visualization:``when I search by visualization, I prefer an exact match and exclude all designs without my selection''.
However, before applying the retrieved designs to their own scenarios, users have to understand the visual encodings. 
Both junior (\workshop{P2, P3}) and senior (\workshop{P10, P11}) Ph.D. students encountered problems understanding retrieved designs. 
\workshop{P2} appreciated the textual explanation about the data types and task types, commenting that ``\textit{the indication of visual encoding helps me to understand the design easily}''. 
They expected more detailed descriptions of the visualizations, not only the types.

\textbf{``Tasks'' are mixed in understanding.} 
Even though two participants ranked tasks first, 
half of the participants ranked it the last.
A common problem is a gap between the original analytical question and low-level tasks. 
\workshop{P12} commented ``\textit{I am fuzzy about the task types, so I prefer to consider how to visualize all the data first}''.
Besides,
\workshop{P10} pointed out that he cannot map tasks such as ``obtaining an overview'' to low-level tasks.
\workshop{P3} explained that he would have a different understanding of the tasks of the question.
Those comments call for continued efforts to classify the tasks in VA systems for searching.

\add{\textbf{Discussion.} Based on the observations above, the trade-offs between data, task, and visualization during participants' search for designs might be related to two factors, including the accessibility of the criteria for searching and the representation power of the indexing approach used in the search engine.
First, participants might focus on the inputs and outputs of the process of designing VA designs, which are available criteria for searching.
Most participants ranked data as their top choice of search condition as data is the most approachable one among data, task, and visualization.
To design a visual analytics system, researchers and designers usually start with data exploration and then consider appropriate design to visualize the data.
On the contrary, participants who are good at designing might turn to the outputs of the design, i.e., visualizations, and opt to ``regress'' their desired design through searching.
Visualizations are graphical representations of the data, which might be the intermediate search condition for the participants.
As mentioned by \workshop{P1},\textit{`` when I saw the column timestamp, location, and prices, I immediately came across a line chart to represent purchases by time and location. Then I searched the designs based on the line chart.''}
In this case, the participant considered the data but chose to use visualization as a representation of the data for searching.
Analytical tasks are also important in deriving designs, but a common problem is the gap between the original analytical question and low-level tasks, which makes tasks uncertain at the early stage of design.
Moreover, an analytical task can be approached through multiple design choices.
For example, designers can use different visualizations in different layouts (e.g., overloaded and mirrored) to compare values~\cite{lyi2020comparative}. 
Therefore, compared to considering tasks at first, practitioners might turn to questions like how to represent the data and what visualizations might be more aesthetically pleasing.}

\add{Second, participants might suffer from an insufficient capability of \system{}-alpha to represent VA designs.
As discussed above, participants turn to visualization instead of data might result from the lack of a more representative method to search visual designs.
Moreover, the tasks were not clear enough so participants chose not to use the task as their first choice to search. 
To help practitioners better retrieve designs and further understand their design preferences, we concluded several requirements that might help improve the representativeness of the \system{}.}

\subsection{Requirements for the Index Design}
\label{sec:requirements}
We derived three key requirements for improving the current index design according to the findings:
\begin{enumerate}[label=\textbf{R\arabic*:}]
    \item \textbf{Integration of data and visualization.} 
    Data and Visualization are the most preferred. All users' comments on data and visualization mention the relations between data and visual channels, namely, visual encodings.
    The indication of visual encodings can help users better understand how the data can be applied to the design. Inspired by the comments, we aim to propose an efficient method to describe the visual encodings in view designs.
    \item \textbf{Description of visualization composition.} Many view designs are composite visualizations, and the composition reflects the data relationship between visual elements. For example, a common VA technique is the ``glyph scatterplot'' where each scatter is represented by a glyph for additional multi-dimension attributes. Such relationships are hard to describe using existing methods. Additional descriptions of such a relationship are required.
    \item \textbf{More detailed descriptions of analytical tasks.} 
    Users' difficulties in mapping real analytical questions to low-level tasks might be because of the lack of analysis goals, such as summarize, compare, and explore~\cite{brehmer2013multi}. Comparing distribution and summarizing distribution might require visualization designs that are different in visual encodings and layouts. Besides, graph-related tasks are not investigated in depth. Therefore, we decided to incorporate additional task taxonomies with multiple levels.
\end{enumerate}
\section{VAID}
\label{sec:structure}

In this section, we introduce the process of index design based on the above requirements.
We intend to represent the index within the JSON structure since the index may include nested elements (\textbf{R1}, \textbf{R2}).
In addition, we elaborated on task characterization using the multi-level typology of VA tasks~\cite{brehmer2013multi} (\textbf{R3}).
Formally, we call the structure ``\system{}'' in the paper, which consists of a two-tuple:
\begin{equation}
    \system{} = Task + Design
\end{equation}
In the upcoming section, we introduce the task and design in detail.

\subsection{\system{} Task}

\begin{figure}[htbp]
    \includegraphics[width=\linewidth]{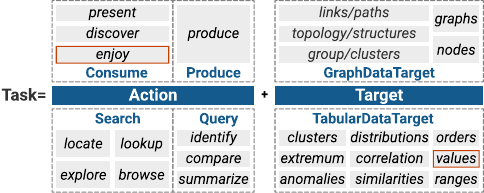}
    \caption{
    A VA task consists of a dual-key index.
    The action's value is selected from four classes, with each class having a single subclass chosen, so as the target's value. The task \textit{``enjoy + values''} is exemplified in red strokes.}
    \label{fig:task}
\end{figure}

Given that the low-level task taxonomy might be insufficient for users to understand VA tasks, we improve the task structure based on Brehmer and Munzner's taxonomy of VA tasks\rr{~\cite{brehmer2013multi}}.
In their taxonomy, a VA task is described with three levels, namely, \textit{why}, \textit{what}, and \textit{how}.
The \textit{why} level refers to the goals of VA, such as present, compare, and browse.
This level also describes human behaviors during analysis, so it is named ``action''~\cite{munzner2014visualization}.
The \textit{what} level explains the analytical ``targets'' in VA, such as raw data, specific attributes, or data patterns.
Our current task taxonomy can be regarded as a subset of these targets.
Moreover, the \textit{how} level describes the methodology used to achieve the ``actions'' and ``targets,'' including view designs and algorithms.
In this work, we focus on the structure of view designs for the \textit{how} level and do not consider it a part of the analytical task structure.

We use action-target pairs to describe the analytical tasks.
The actions are the same as their definitions in the original taxonomy.
For the targets, we refer to Amar et al.'s low-level task taxonomies for tabular data~\cite{amar2005low} and Lee et al.'s task taxonomies for graph data~\cite{lee2006task}.
The classifications of actions and targets are summarized in Fig.~\ref{fig:task}.
\rrindex{S3.1}\add{
The detailed annotation process for tasks is presented in the subsequent subsection, conducted together with the VAID design.
We identified action-target pairs only when explicitly mentioned by the original authors, and any disagreements during the process were resolved following the same strategy.
}

\subsection{\system{} Design}
\label{sec:designspec}
For view designs, we aim to identify visual encodings inside, which are the mappings from data to visual channels and layouts.
Specifically, we regard each design as a composite visualization~\cite{javed2012exploring, deng2022revisiting}.
We begin by recognizing the overall layout, such as faceting, and then break it down into various visual components.
Each component is an independent visualization of specific types, such as bar charts, line charts, and Sankey diagrams.
For the designs of well-crafted glyphs that are not just combinations of different visualization types, we regard them as ``others'' type.
Then we recognize the visual encodings for each component.

We use Vega-Lite~\cite{satyanarayan2016vega} as a starting structure because their JSON syntax is intuitive for representing visual structures.
Specifically, it characterizes a visualization with the fields of ``mark'' and ``encoding''.
In the field ``encoding'', the data ``field'', ``type'', and ``aggregate'' are further specified.
Moreover, it supports basic visual compositions, such as faceting, concatenating, and layering. 
We iteratively developed the structure to cover the collected view designs and annotated each design.
The process of extension and annotation consisted of the following four stages. 

\begin{figure*}[!htb]
    \includegraphics[width=\linewidth]{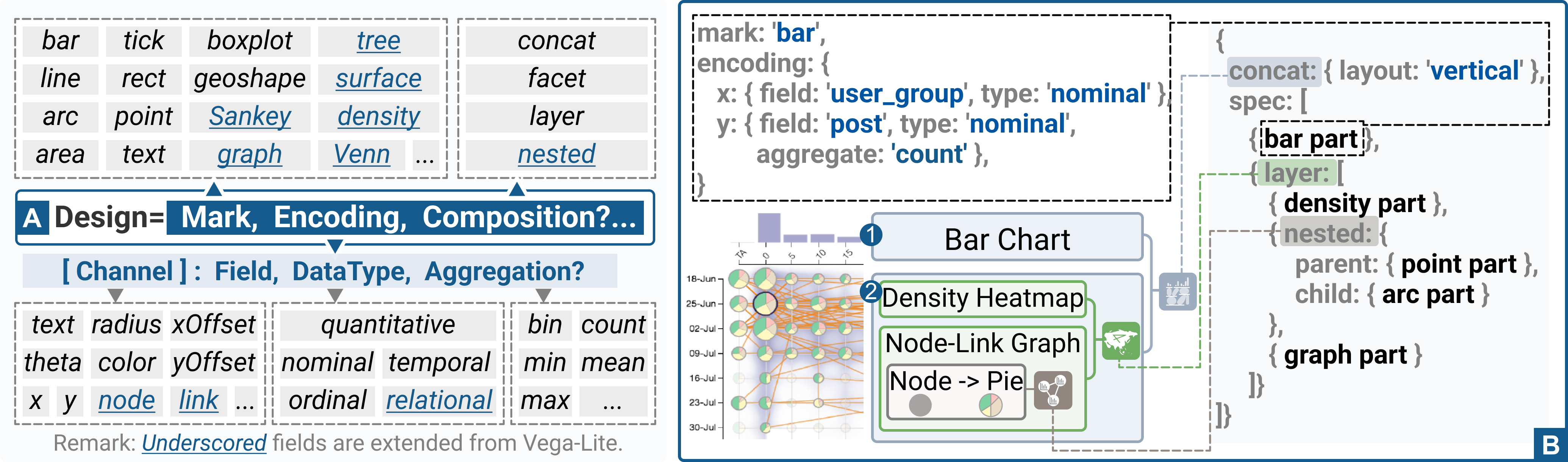}
    \caption{We follow a Vega-Lite style to describe complex view designs. (A) Our formal index structure for visual designs and (B) an example with the index. 
    The structure for representing each chart, including marks and encodings, is simplified and indicated using ``[] part'' with black text. To illustrate, we provide an example of the bar part in the upper left corner.}
    \label{fig:spec_example}
\end{figure*}

\begin{figure*}[!htb]
    \includegraphics[width=\linewidth]{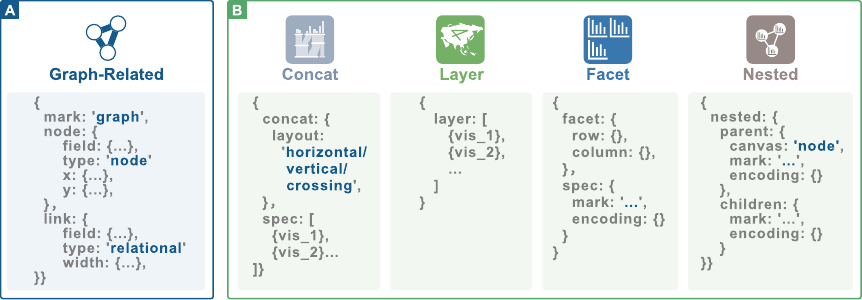}
    \caption{Example indexes of (A) graph-related visualizations and (B) visualizations with different compositions.}
    \label{fig:data_structure}
\end{figure*}

In the first stage, four authors annotated visualizations with original Vega-Lite. 
We discovered that the Vega-Lite did not support the description of graph-related visualizations, such as Sankey diagrams and tree visualizations, which are common visualization types in view designs. In addition, complex visual compositions are not supported, such as embedding glyphs in graph nodes.
We attempted to extend the structure based on the failed cases.
We extended the original Vega-Lite structure from three perspectives:
\begin{itemize}
    \item First, we added additional data types (e.g., relational data) and regarded ``node'' and ``link'' to be the two visual channels of graph-related visualizations.
We further specified the properties of the nodes and links, such as positions and widths, under the ``node'' and ``link'' labels.
An example structure is presented in Fig.~\ref{fig:data_structure}A.
    \item Second, to handle complex visual compositions, such as embedding glyphs in graph nodes~\cite{elmqvist2009hierarchical}, we have added a composition type ``nested''.
The nested visualizations are represented by specifying the ``parent'' and ``children'' components.
We also use a key ``canvas'' to indicate which elements of the parent are the embedded children components.
The structures of the compositions, i.e., ``concat,'' ``layer,'' and ``facet,'' and ``nested''  composition, are shown in Fig.~\ref{fig:data_structure}B.
    \item Third, the mark types supported by Vega-Lite are also insufficient for the representation of view designs.
We have extended the mark types by adding new ones like graph, Sankey, and radar, referring to the typologies proposed by Borkin et al.~\cite{borkin2013makes}.
It is noted that the newly added mark types are not graphical primitives that are elemental building blocks of the visualization.
Instead, some of them are ``\textit{macros for complex layered graphics that contain multiple primitive marks}''\cite{satyanarayan2016vega}, which are consistent with the definitions of Vega-Lite.
Fig.~\ref{fig:data_structure} provides an example of complex composition relationships, such as nesting bar charts into node-link graphs.
\end{itemize}

In the second stage, 
\rrindex{S3.2}\rr{
we independently annotated the views using our labeling system, including tasks and designs, according to the descriptions in the ``Visual Design'' and ``Case Study'' sections in the original papers.
In instances where the desired information was unavailable, we reviewed the entire paper.
Additionally, for the design structure, we identified cases not addressed by the extended structures.
Weekly online discussions are conducted to synchronize and address cases. 
The procedure includes initially creating a shared document that outlines the structure and subsequently updating it until all cases can be covered with the extended structures.
A final document incorporating design structure and annotation examples for various cases was derived at that stage.
In the third stage, each author revised their annotation results using the document from the second stage. 
One of the authors systematically compared all results, labeling any disagreements. 
These conflicts were recorded in our system, and resolution occurred through discussions among all authors during weekly online meetings, resulting in updated results directly.
}
Finally, one of the authors double-checked the results again for all details.
As a result, we obtained the index structure for \system{} design (Fig.~\ref{fig:spec_example}) and annotated 442 view designs following the structure.

During the annotation, we followed the idea of consistency.
In detail, we annotated the view while striving to preserve the original Vega-Lite structure as much as possible.
Compared to the original Vega-Lite structure, we extensively expand the properties of composition, marks, encoding types, and data types based on the VA designs we collected.
It is noted that the Vega-Lite structure also provides powerful operators for data transformations, such as filtering.
However, in VA research, many techniques involve complex data processing methods, such as dimensional reduction, and the classification and identification of these methods are challenging.
In this work, we currently focus on view designs and only use part of the Vega-Lite structure to characterize visual encodings (e.g., excluding style-related parameters) that help for better view indexing and understanding.

Moreover, we followed the idea of minimization, i.e., choosing the one with the least number of duplications, to address the problem when there are multiple solutions to a visualization.
For example, the component in Fig.~\ref{fig:spec_example}(B2) can be regarded as ``a layered visualization with a density plot and a graph'' and ``a facet visualization whose elements are pie charts'' from the perspective of implementation.
The positions of the graph nodes and pies both repeatedly encode the row and column attributes.
Referring to the original descriptions~\cite{fu2017mooc}, the pies without links would fade out, indicating a one-to-one mapping of the graph nodes and pies.
Therefore, it would be more appropriate to consider the visualization as a layered one composited by a density plot and a nested graph visualization (pie charts embedded into the nodes).
For another example, a faceted visualization can be considered as a concatenation of a list of similar visual components.
Representing the chart with ``concat'' has to duplicate the structures of similar visual components multiple times.
Instead, it is much neater and more accurate to use ``facet'' to represent it in the context of data visualization.
\begin{figure*}[!htb]
    \includegraphics{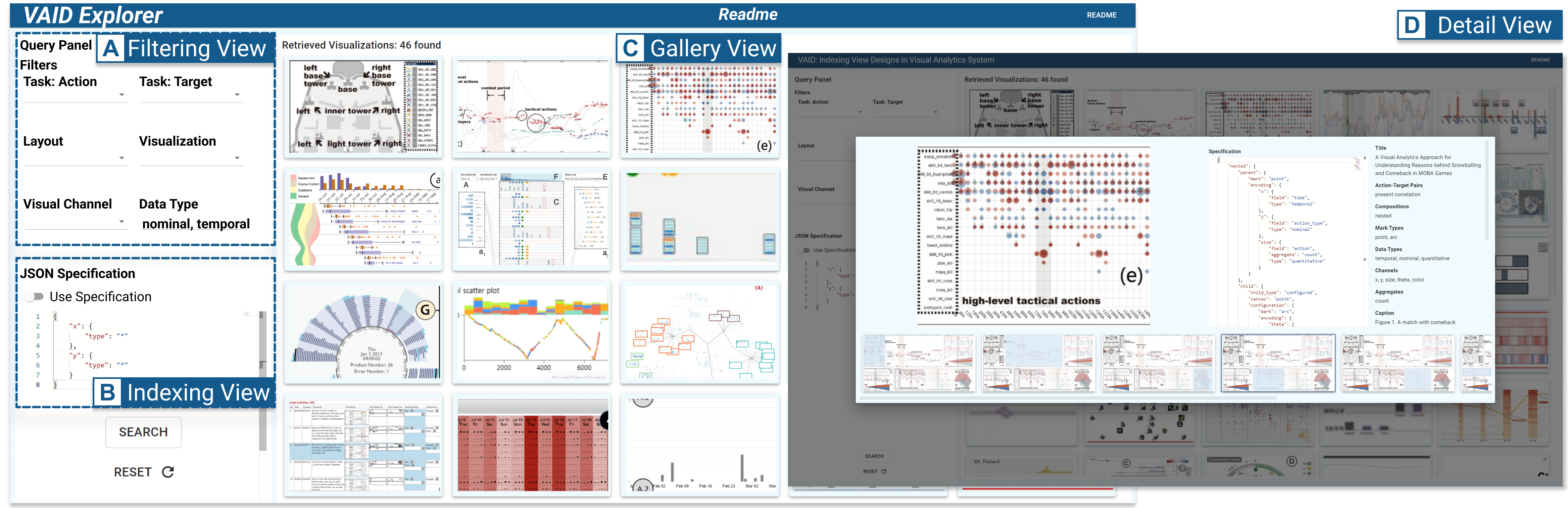}
    \caption{The \system{} Explorer contains a filtering view (A), an indexing view (B), a gallery view (C), and a detail view (D). }
    \label{fig:prototype}
\end{figure*}

\section{Evaluating VAID through question-based user study}
\label{sec:study}

\rrindex{S2.2}\rr{We conduct a user study to evaluate if \system{} can assist users in view design.}
To allow users to experience the design search using \system{}, we developed a prototype system named VAID Explorer.
In this section, we first provide a brief overview of the prototype, followed by an in-depth discussion of the user study.

\subsection{\system{} Explorer}
We will present the prototype and describe how it is used in the following section.

The prototype includes a filtering panel (Fig.~\ref{fig:prototype}A), a gallery view (Fig.~\ref{fig:prototype}C), an indexing view (Fig.~\ref{fig:prototype}B), and a detail view (Fig.~\ref{fig:prototype}D).
The filtering panel (Fig.~\ref{fig:prototype}A) supports view design search.
Users can select values corresponding to different keys introduced in Sec.\ref{sec:structure}. 
We also develop an indexing view (Fig.~\ref{fig:prototype}B) that enables users to input structural indexes with JSON syntax.
\rrindex{S3.3.2}The retrieved results will be displayed in the gallery view (Fig.~\ref{fig:prototype}C).
When clicking on a result, a detail view (Fig.~\ref{fig:prototype}D) will pop up, \rr{showing the VAID along with other contextual metadata (e.g., paper title, paper keyword, figure caption). }
Users can also explore the designs of other views.

\begin{figure}[!htb]
    \includegraphics{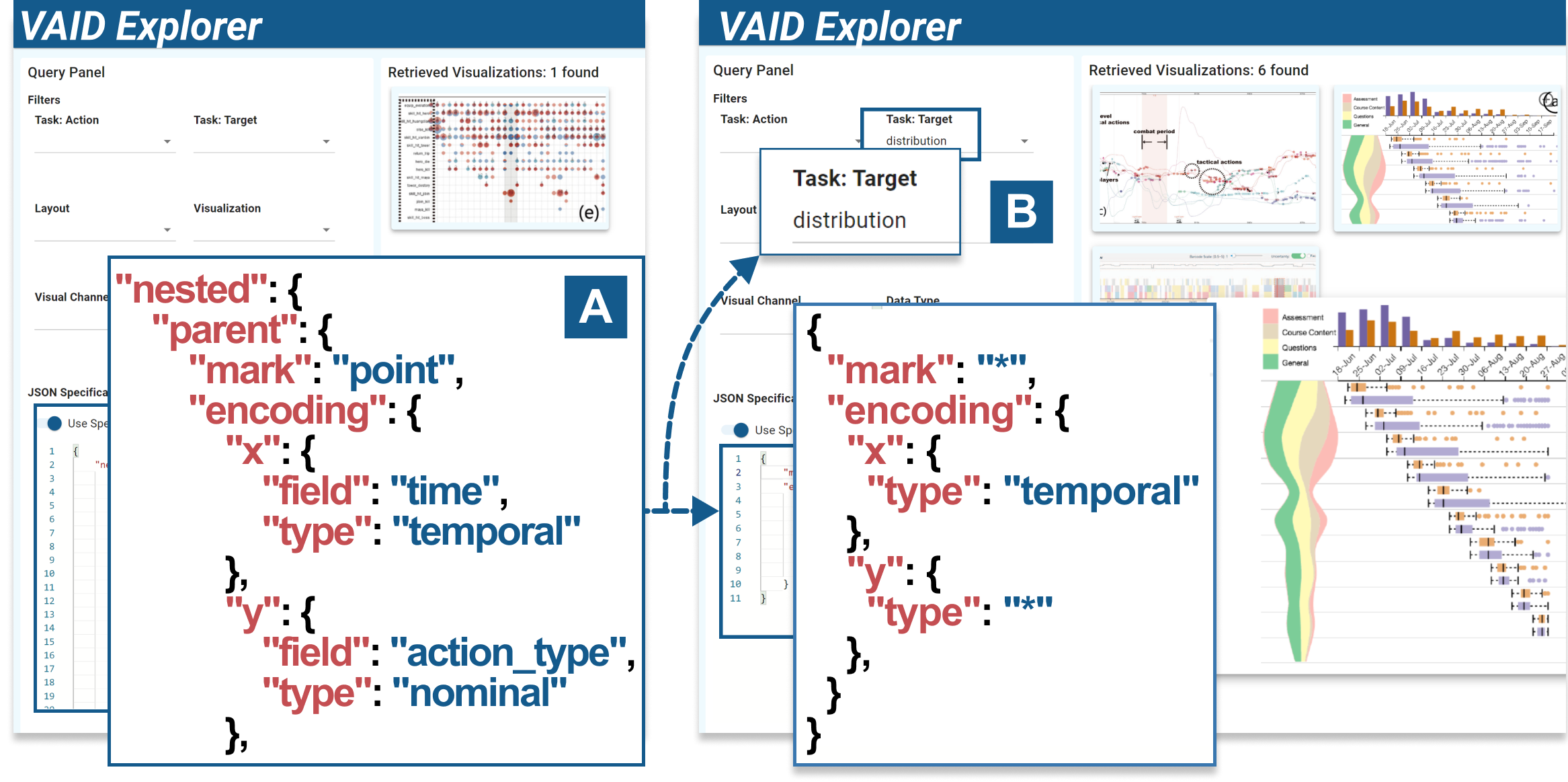}
    \caption{A usage scenario using \system{} to design new visualizations for visual analytics. 
    The prototype supports searching view designs by detailed indexes (A) and key values (B).}
    \label{fig:scenario}
\end{figure}

We present a concise scenario featuring Sherry, a visualization researcher in the field of urban planning, to demonstrate the usage of the prototype.
She was given a dataset about credit card records, where there are four columns of ``card id,'' ``time,'' ``store,'' and ``item name.''
She wants to identify the most popular purchasing time and store, but she's uncertain about how to represent this data.
With the \system{} Explorer, she first starts from the filtering panel (Fig.~\ref{fig:prototype}(A)) and selects two data types, nominal and temporal.
Retrieving 46 view designs, she explores the results and discovers that the third example (Fig.~\ref{fig:prototype}D) can show the data across ``time'' and ``store.''
To further show the popularity of different times and stores, she wonders how to show distribution with a similar design.
She copies the index of this design into the indexing view (Fig.~\ref{fig:scenario}A).
She revises and only keeps the sub-structure of the index, and selects the target of ``distribution'' (Fig.~\ref{fig:scenario}B).
She identifies a design with bar charts and area charts showing the summarization of the nominal dimension and temporal dimension, respectively.
Based on the example, she has some preliminary thoughts on the view design.

\subsection{Study Setup}
\rrindex{S2.2}\rr{Our goal is to understand if participants can understand the retrieved designs (e.g., visual encodings) using \system{} and use the prototype to obtain design inspirations for VA problems.}
Specifically, we ask participants to search for visualizations to solve six well-designed VA problems \add{and subsequently, design views.}
The VA problems have varying complexity and are related to specific analytical tasks, mark types, composition types, or data types,
such as ``to find visualizations that encode a three-dimensional dataset with a nominal, a quantitative, and a temporal field'' and ``to find VA designs for comparing distributions''. 
\add{For the design task, we choose Mini-Challenge 2 from IEEE VAST Challenge 2022~\cite{vastchallenge2022}.}
The detailed list can be found in the supplementary material.
Given the retrieved results, participants were asked to explore the results and select one visualization of interest for in-depth investigation, that is, to understand the design by reading images, indexes, and other metadata (e.g., titles and captions).
Lastly, we ask participants to explain the \rr{design} of the visualizations.

\textbf{Participants.} We recruited 12 visualization practitioners (\userstudy{\userstudy{U1}-\userstudy{U12}}) \rrindex{S3.1.3}\rr{from our institution through social media and word-of-mouth} who reported having experience in creating visualizations for data analysis. The participants include 5 females and 7 males with various backgrounds, including \rr{computer science, urban, and digital media design}. 
They are used to analyze data with toolkits such as Python, R, and MATLAB for data analysis. In addition, the libraries they use for data visualizations are \add{Vega-Lite, }Excel, Python Matplotlib, and Javascript D3.
The participants in this user study all have adequate data visualization or design knowledge but have varying expertise in designing more complex VA systems.

\textbf{Procedure.} 
\rrindex{S3.1.3}\add{All studies were conducted through one-on-one online meetings.}
Each study consisted of two sessions: a training session (15 minutes) and an experiment session (20 minutes). In the training session, we introduced the definitions of \system{}, including taxonomies of task and design.
Then we introduced the use of the prototype. All participants were allowed to use and explore the data freely to get familiar with the prototype.
In the experiment session, each participant was asked to accomplish \add{seven questions (or tasks, but to avoid confusion with the ``task'' dimension in \system{}, we use the term question here). }
During the whole study, we followed the think-aloud protocol. Participants were requested to speak out about \rrindex{S2.1.2}\rr{their understanding of the retrieved view design} and their thoughts about the \system{} or prototype when accomplishing tasks.
The study ended with a post-study interview session as well as a questionnaire for rating the \system{} from different dimensions. 
\add{The whole user study lasted about 1-1.5 hours. Each participant received \$9 as compensation.
 The authors took notes to record feedback during the study.}
\begin{figure}[!b]
    \includegraphics[width=\linewidth]{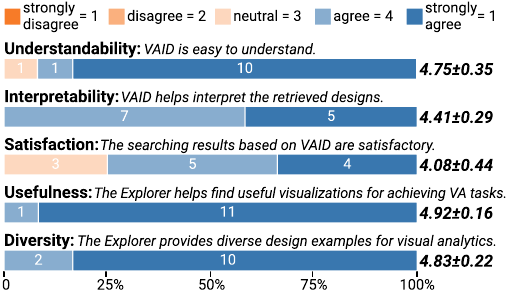}
    \caption{User ratings from the perspectives of understandability, interpretability, satisfaction, usefulness, and diversity. The number on the right illustrates the average score and the 95\% confidence interval.}
    \label{fig:search-userstudy}
\end{figure}
\subsection{Results and Feedback} 
\rrindex{S1.2.2}\add{All users successfully found the required designs and comprehended the design using VAID. For Question 7,  they designed and sketched several views. All sketches can be found in the supplementary materials.}
The quantitative results from the participants were very positive, as shown in Fig.~\ref{fig:search-userstudy}. Eleven out of twelve participants strongly agreed that based on VAID, the prototype helps to find useful visualization designs for achieving the VA tasks in the study. Similarly, 10 participants strongly appreciated the diversity of VA designs in \system{} Explorer. 

\rrindex{S1.2.1}\add{\textbf{VAID is easy to understand and benefits design understanding.}
Most users (11/12) find VAID easy to understand. 
Some (\userstudy{U1}) attribute this ease to their familiarity with Vega-Lite, facilitating a swift adaptation. 
Others, unfamiliar with Vega-Lite, emphasize the significance of VAID's JSON format and declarative language for clarity. 
\userstudy{U6} further emphasized that basic chart knowledge aids in understanding VAID's design structure.
Additionally, users appreciate VAID's role in simplifying visual encoding comprehension. 
\userstudy{U5} and \userstudy{U7} noted that VAID complements textual elements like titles and captions, offering insights beyond what these elements convey alone. 
\userstudy{U1} underscored VAID's importance in clarifying glyph-related aspects to prevent uncertainties in visual interpretation. 
Their viewpoint confirmed the fulfillment of design requirement \textbf{R1}.
Despite its clarity, \userstudy{U7} still mentioned that referencing the research paper may still be necessary for more complex aspects.}

\add{\textbf{VAID Explorer enables users to swiftly derive initial designs based on the given question.}
All users start Question 7 very quickly.
For example, \userstudy{U12} started by selecting different combinations of ``actions'' and ``targets'' values that might conform to the questions and said, ``\textit{Previously, I required time to grasp background information; however, now I can randomly select filter parameters to explore potential designs. Examining these designs helps me better understand the question and formulate a vague initial design as a starting point.}''
Users agreed that the structure of VAID aligns with common design strategies, wherein they typically approach the view design by considering \textit{data, task, and visualization}. 
\userstudy{U11}commented ``\textit{The filter options align with my way of thinking about the design question. I find it easy to kickstart the process, given that the question description provides the necessary information. Subsequently, I can quickly discover inspiration.}''}

\add{\textbf{VAID facilitates comprehensive search.} 
The majority of participants expressed satisfaction with the search results during the exploration.
Users pointed out, ``\textit{While exploring, I aim to retrieve all pertinent designs without any omissions.}''
\userstudy{U6} complimented, ``\textit{My personal preferences may introduce biases and potentially result in overlooking valuable papers. The use of the system, however, ensures a more thorough exploration.}''
Additionally, \userstudy{U4} highlighted, ``\textit{It operates as a knowledge-based retrieval system, effectively supplementing my knowledge.}''
Compared to the preliminary study, the extension of VAID facilitates more flexible search options.
Specifically, \userstudy{U3} and \userstudy{U5} appreciated the flexible task options when designing a multi-view VA system. They emphasize its effectiveness within a VA system, where the task's target remains constant while actions vary between views.
\userstudy{U5} also commented with an example, `\textit{`VAID enables designing VA systems in a manner of progressive exploration, identification, and localization of anomalies.}''
Their opinion verified that the design requirement \textbf{R3} had been fulfilled.}
\highlight{However, even with the enhanced flexibility in search options provided by VAID, some users still faced challenges while choosing actions.
U11 expressed that the concise one-word descriptions lack intuitiveness. 
She suggests,  ``\textit{including examples and images as hints would help me make more informed task choices.}''
With the development of large language models (LLMs), a potential solution is to introduce an LLM to help translate the analytic questions into abstract actions and targets, which might lower the burden of using the system.
}

\add{Moreover, the search function was praised by users for its user-friendly nature, as \userstudy{U11} said, ``\textit{The filtering and indexing view can complement each other. While the filtering is easy to use but less precise, the index search can assist in retrieving more specific designs, like a bar chart with temporal data on the x-axis.}''
\highlight{Despite the positive feedback, there is room
 for improvement.
The current combination of multiple options may result in limited or no results, potentially leading to neutral satisfaction among some users as shown in Fig.~\ref{fig:search-userstudy}.
\userstudy{U12} recommended incorporating a partial matching mechanism to guarantee a significant number of retrieved designs, even under slightly stringent conditions.
\userstudy{U8} echoed similar sentiments, proposing the addition of a recommendation mechanism akin to common search engines.}
Additionally, \userstudy{U5} expressed a desire for improved linkage between the filtering view and the indexing view. He mentioned, ``\textit{Typing a JSON structure from scratch is not easy, but editing one is simpler.}'' \userstudy{U5} hoped for an initial draft in the index view after selecting options in the filtering view.
Therefore, valid ranking mechanisms and linkage query functions for view designs can be developed in the future to support an effective visualization query system.}

\add{\textbf{VAID facilitates incremental design, helping users refine their designs step by step.}
View designs are mainly composite visualizations~\cite{deng2022revisiting}, which can be further broken down into various basic charts (Sec.~\ref{sec:designspec}). Designers usually start from one basic visualization type that can be decided. 
For example, many users recognize ``maps'' in Question 7 due to the urban scenario. 
Some users also search for initial designs by integrating data, tasks, and their expertise.
Building upon this concept, the VAID Explorer empowers users to employ basic visualizations as search parameters, enabling the design of intricate and extended designs (\textbf{R2}).
\userstudy{U10} stated, ``\textit{I usually use several basic charts to meet design requirements. After that, I explore ways to refine the design by integrating these charts into a unified view.}'' She believed that using our tool makes this refinement step easier than before.
\userstudy{U9} conveyed a strong appreciation for layouts, citing challenges in keyword-based searches due to reliance on personal knowledge, ``\textit{VAID Explorer addresses this by offering nested or layered layout options}''. 
This preference aligns with sentiments from \userstudy{U5} and \userstudy{U7}, who stress spatial constraints in VA views and the importance of thoughtful layout choices within limited space. 
Meanwhile, some users also suggest that the complexity of the design allows for a selective approach to the retrieved designs. For example, in Question 7.3, \userstudy{U7} may opt to utilize only the color and size encoding within a view design from TPFlow~\cite{liu2019tpflowa}.
\userstudy{U5} also employed a similar approach. Initially, he selected the circular bar design from one view in \cite{zhou2019visual}, and later chose the area chart design from one view in \cite{zhao2017annotation} to address the question.
}

\add{\textbf{VAID enhances design aesthetics.}
VAID not only aids in completing a design but also provides additional assistance, as highlighted by \userstudy{U2}, who pointed out that VAID contributes to enhancing aesthetics during view design.
Moreover, users may further improve the design using VAID, even when it effectively achieves the intended task.
For example, in Question 7.1, \userstudy{U9} initially obtained a design that satisfied the question. However, upon observing a particular view in Volia~\cite{volia2018cao}, she recognized the potential of using quadrilaterals or hexagons as the smallest units during map segmentation, which helped her to refine the design accordingly.
In Question 7.3, \userstudy{U6} noticed that bar charts were commonly used in previous designs, leading to boring designs that lack adequate novelty. This prompted \userstudy{U6} to explore alternative views within the last referred VA interface, aiming for more varied designs. By combining this exploration with insights from the VAID structure, \userstudy{U6} took a radical approach.}

\section{The analysis of VA design collection using VAID}
\label{sec:analysis}
After validating the usefulness of VAID, we explore the ``design demographics''~\cite{hoque2019searching, battle2018beagle} based on \system{} for view designs \rrindex{S3.1.3}\add{with collected views in Sec.~\ref{sec: data preparation}}, which demonstrates VAID enables fine-grained exploration and understanding of existing view designs.
In particular, we report on the statistics, including the frequency and co-occurrence patterns for the analytical tasks and visual designs.

\subsection{Overview}
We first analyze the indexes to understand the composition of visualization designs in visual analytics.

\begin{figure}[!htb]
    \includegraphics[width=\linewidth]{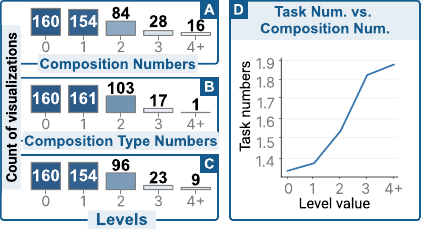}
    \caption{Overview of the view designs in terms of compositions.}
    \label{fig:data-overview}
\end{figure}

\textbf{Only 38\% view designs can be implemented with Vega-Lite.}
We investigated the indexes and opted to understand whether these visualizations can be implemented with common compositions (layer, concat, and facet) of basic mark types with Vega-Lite.
We discovered that only 38.2\% (169/442) designs could be specified with pure Vega-Lite structures.
The results indicate that researchers tend to use novel techniques in VA systems to visualize the data, which demonstrates the unique value of our structures.
The limitations of Vega-Lite mainly lie in the limited mark types and the lack of support for graph-related data and nested visualizations.
The limited expressiveness of declarative visualization grammars may be a reason for the result that the visualization designs of most existing VA systems are implemented with lower-level Javascript libraries (e.g., D3.js), as they could provide flexible customizations for the designs.

\textbf{About 64\% view designs are composite visualizations.} 
Composite visualizations combine multiple visual components together along specific directions (e.g., ``layer'', ``concat'' and ``facet'' in Vega-Lite) or in a hierarchical manner (i.e.,  ``nested''), which can make the visual components well-organized and easy to interpret.
As shown in Fig.~\ref{fig:data-overview}A, 63.8\% (282/442) of the view designs contain composite visualizations. 
We focus on the number of composition labels in a structure.
Fig.~\ref{fig:data-overview}A shows that 34.8\% (154/442) of visualizations contain only one composition.
3.6\% (16/442) visualizations contain at least four compositions. 
A visualization can have several types of compositions (four different types in total).
Only one visualization has used the maximum number of different composition types, which is four (Fig.~\ref{fig:data-overview}B). 
Fig.~\ref{fig:data-overview}B shows that most visualizations have no more than two composition types.

\textbf{90\% composite visualizations have a hierarchy level $\mathbf{\leq}$ 2.}
As described in \autoref{sec:designspec}, the visualization is represented in a hierarchical JSON syntax with multiple levels of composition. For example, the visualization presented in Fig.~\ref{fig:spec_example} has a composition level of three.
For composite visualizations, 54.6\% (154/282) have a level of one, 34.0\% (96/282) have a level of two, and 8.1\% (23/282) have a level of three.
Only 9 visualizations have a level of four, which is the maximum level of the hierarchy.
The numbers indicate that most composite visualizations only use one or two levels of composition.
Adding more levels of composition demands encoding more data columns, which might go beyond the requirement of analysis scenarios. 
Moreover, more levels of composition increase implementation difficulty and visual complexity.

\textbf{More compositions, more tasks achieved.} Investigating \system{}, we discover that all visualizations have at least one action-target task. The ones with composition achieve 1.49 tasks on average, while the ones without composition are designed for 1.33 tasks on average.
Fig.~\ref{fig:data-overview}D shows the average number of tasks vs. the number of compositions.
Overall, the number of tasks to be solved will increase with the increase of compositions.

\subsection{Frequency Analysis}
\begin{figure*}[!ht]
    \includegraphics[width=\linewidth]{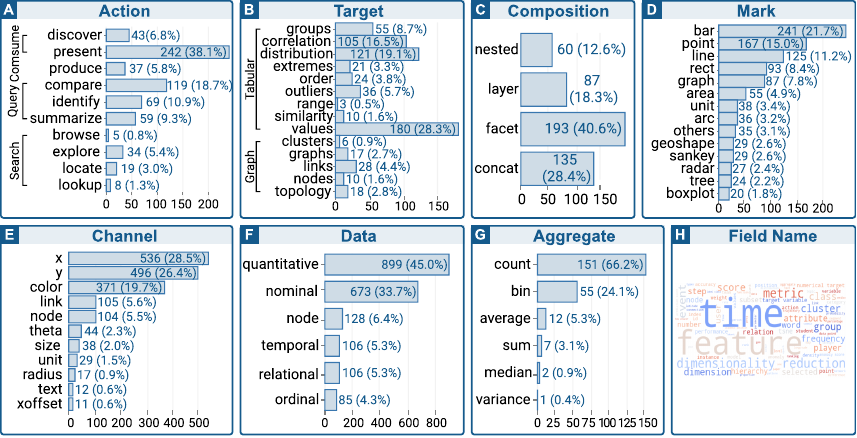}
    \caption{The property distribution of view designs: actions (A), targets (B), composition types (C), mark types (D), visual channels (E), data types (F), aggregate types (G), and field names (H).}
    \label{fig:statistics}
\end{figure*}
We then report on the frequency of different property values and findings in \system{}.

\textbf{Actions: ``low-level'' actions are the most used analytic actions.} 
As shown in Fig.~\ref{fig:statistics}A, \textit{present} is the most frequent action.
The results show that many designs are merely used for data exhibition.
Therefore, we excluded \textit{present} in the later analysis since designers commonly use terms like ``show,'' ``visualize,'' and other ambiguous words to explain their use of such view designs.
After excluding this category, \textit{compare}, \textit{identify}, and \textit{summarize} are the most popular.
These three actions are categorized into ``low-level'' query actions by Brehmer and Munzner~\cite{brehmer2013multi}.
For the goal of searching, \textit{explore} is the most popular one, which stands for exploratory analysis.
\rrindex{S3.1.2}\rr{Interestingly, we did not discover designs that are used for \textit{enjoy}, showing the difference between visual analytics and infographics.}

\textbf{Targets: domain-specific values are the most popular targets.} 
From target distribution (Fig.~\ref{fig:statistics}B), we discovered that the most popular target is \textit{value}, which refers to visualizing values that are computed from metrics or algorithms.
This reflects the features of VA, which closely collaborates with domains and utilizes data mining techniques for data preprocessing.
\textit{Distribution} and \textit{correlation} are the second and third popular targets.
The results demonstrate that understanding the correlations and distributions of the attributes are the key indicators for data patterns in VA.
For graph data, \textit{links} and the whole \textit{graphs} are the most frequently visualized targets.

\textbf{Compositions: simple is preferable.} 
Composition distribution is presented in Fig.~\ref{fig:statistics}C. \textit{Facet}, which organizes visualizations of the same types by rows and columns, is the most used composition type in VA. 
This simple composition conveniently visualizes one or two more dimensions with simple visualization building blocks (e.g., scatterplot matrix).
\textit{Concat} is the second popular type, which refers to placing visualizations with different types side by side. 
\textit{Nested} composition is not covered by the original Vega-Lite. 
Although it has the smallest proportion, it accounts for more than 10\%.

\textbf{Marks: basic types dominate.} 
In a composite visualization, each visual component is regarded as a specific mark type. We display the mark types that have more than 20 records (Fig.~\ref{fig:statistics}D). The distribution demonstrates that \textit{bar}, \textit{point}, \textit{line}, and \textit{rect} are the most popular mark types, which are also basic mark types in Vega-Lite. For the types that Vega-Lite does not cover, \textit{graph} and \textit{unit}~\cite{park2017atom} visualizations rank $5^{th}$ and $7^{th}$ among all types. The type \textit{others} ranks $9^{th}$, indicating that glyph visualizations are also commonly used in view designs.

\textbf{Channels: most relate to the Cartesian coordinate system.} We show the visual channels with more than 10 records in Fig.~\ref{fig:statistics}E. The channels \textit{x}, \textit{y}, and \textit{color} are the most popular. The channels \textit{link} and \textit{node} are also frequently used because of the graph-related visualizations, such as Sankey diagrams, graphs, and trees.

\textbf{Data Types: about 90\% fields are quantitative and nominal.} Among all data types, \textit{quantitative} data and \textit{nominal} data are the most frequently encoded in the visualizations (Fig.~\ref{fig:statistics}F), followed by \textit{node} data, which is commonly used in graph data.

\textbf{Aggregate: binning/counting, or more complex operations.} Aggregate types are labeled based on Vega-Lite aggregation operations (Fig.~\ref{fig:statistics}G). Aggregate \textit{count} and \textit{bin} are the most frequently used types, which are usually because of the visualization of histograms. Other aggregate types are not frequently used in view designs. The reason might be that complex data processing methods and metrics are adopted in VA, instead of basic aggregation strategies, such as \textit{sum}, \textit{median}, and \textit{variance}.

\textbf{Field Names: time and feature analysis are main characters.} Field names are the terms used in the original VA research papers describing the fields. We show the word frequency of field names using word cloud (Fig.~\ref{fig:statistics}H). From the word cloud, we immediately discover that the words \textit{time}, \textit{feature}, \textit{metric}, and \textit{dimensionality reduction} have a relatively large size, indicating the values about these terms are frequently used in VA research.
\section{Discussion}
In this section, we discuss the potential avenues for future research on VAID and its limitations.

\subsection{Opportunties for Future Research}
We identified multiple research opportunities grouped into three primary avenues.

\textbf{First, VAID offers the potential to enhance view design assessment.}
While our statistical analysis of 442 designs has yielded valuable insights (Sec.~\ref{sec:analysis}), there remains an opportunity for deeper analysis through the integration of VAID.
For example, in the current process of VA design, the selection of design alternatives is mainly guided by design principles~\cite{wu2022defence}. 
VAID can retrieve potentially useful designs regarding data and tasks, which complements design alternatives for a more comprehensive discussion and justifications.
Such an index structure accompanied by a database can help to improve the rigor of VA research.
Future research can focus on developing an evaluation method for VA designs based on VAID since this structured approach transforms abstract designs into a more analyzable format, enabling the application of various analysis techniques (e.g., regression, clustering).

\textbf{Second, VAID presents the opportunity to simplify comparisons of view designs.}
Although VA designs have long been criticized for their over-crafted designs for specific domain problems~\cite{wu2022defence}, they might share similarities in specific views, components, and tasks.
As highlighted by \userstudy{U5} and \userstudy{U9}, the importance of comparing designs cannot be understated during the exploration process. 
VAID allows for comparisons in different dimensions, revealing both commonalities and differences in analytical tasks and visual designs. Future research can focus on enhancing the effectiveness of comparisons between different designs.

\textbf{Thirdly, we envision VAID as an initial step toward enhancing the automation in VA.}
While there have been efforts to automate the creation of visualizations~\cite{wu2021ai4vis}, there has been limited exploration of automation within the realm of complex VA design.
Automating VA design requires large-scale datasets in need of training, which necessitates detailed information for VA designs. 
One challenge in this regard is the mismatch between the intensive visual information conveyed and the limited accessible information through captions and figures. In this regard, VAID takes on a crucial role as an initial step in augmenting the accessible information.
Additionally, we encourage the open-sourcing of more VA systems, as they represent valuable outcomes of iterative design. The designs and system code shared through open-source projects will serve as valuable resources for the community. 
Through these collective efforts, we can gradually simplify the production process of VA systems, ultimately achieving automation in VA design.

\subsection{Limitations}
As a first attempt to construct an index structure for VA design from the perspectives of tasks and visual designs, our work has several limitations that warrant future research.

\textbf{Interactions.} 
Interactions are important features of VA systems.
However, it is difficult to recognize the interactions from VA designs, even in an entirely manual manner, since not all interactions within/between views are introduced in the original papers. 
Moreover, the use of static images hiders our analysis of configurable VA systems (e.g., Turkay et al.~\cite{turkay2016designing}), as the configuration frameworks are not reflected in the images.
Facilitating better analysis of view relationships requires parsing live VA systems and constructing the data flow between views.

\rrindex{S4}\rrindex{S3.1.2}\add{\textbf{Generalizability.}
In this study, we designed and evaluated VAID based on high-quality VA designs from top-tier conference papers.
These designs make up a corpus that comprises composite and multiple-view visualizations, which were recognized to be complex and hard to understand~\cite{wu2022defence}.
As a result, VAID is capable of representing visualization designs with complex structures.
We believe that VAID can be used to index and represent a wider range of visualization designs, such as infographics, which are usually facilitated with novel layouts and glyphs~\cite{ying2022glyphcreatora} that improve the expressiveness of information.
However, it introduces additional challenges because infographics usually contain distorted graphical elements for metaphoric representation~\cite{ying2022metaglyph} and the combination of additional modalities, such as text and images.
In this study, we started from the VA community and derived VAID as a kickoff for the research of indexing such complex visualizations.
It requires future research to evaluate and extend VAID with a more general dataset.
}

\textbf{Evaluation.}  
In our two in-lab studies, participants are required to complete the VAST mini-challenge in a short time.
We hope to synthesize the scenario of creating VA designs, but real-world VA design often involves collaboration with domain experts.
While we tried to avoid tasks requiring specialized knowledge, fully simulating authentic collaborative scenarios remains a challenge. 
In the future, we hope to carry on a field study with VAID, asking VA experts to use VAID in their routine design process, observing their behaviors, and gathering more comprehensive feedback from their experience.

\textbf{Scalability.} In this work, the scalability of annotation is limited because it requires extensive visualization knowledge for annotating such a fine-grained structure.
Our work is rooted in the fact that there lack of practical rules and guidance in decomposing view designs in VA.
As a starting point, we manually annotate and fine-tune the structure iteratively with a workshop study, aiming to construct a solid foundation for the indexing.
Such manual efforts were expensive and resulted in a relatively small dataset size.
In the future, we plan to improve \system{} with a combination of machine learning methods.
\rrindex{S3.3.1}\add{
These methods not only ease the effort of manual labeling but also enhance the information available.
In terms of the former, approaches like VisImages~\cite{deng2023visimages} leverage computer vision models to detect view locations in research papers. Efforts can also be made to extract visual structures such as maps~\cite{poco2018bitmap}, charts~\cite{poco2017reverse, savva2011revision, ying2023reviving}, and PowerPoint slides~\cite{shi2022reverse}.
It is possible to adopt deep learning models to detect the positions of visual elements and reconstruct their relations.
Regarding the latter, additional information can be valuable. For instance, utilizing computer vision models to derive color palettes aids in analyzing the emotional tone of designs~\cite{lan2023affective} and inspires future designers~\cite{shi2023destijl}. Annotations and other text information extracted using OCR techniques~\cite{memon2020handwritten} from charts can serve as supplementary material, aiding users in understanding essential information such as the data narrative and context~\cite{ren2017chartaccent}.
}
\section{conclusion}
We built an index structure, \system{}, from visual analytics research papers.
The structure features an index for describing complex VA designs from the perspectives of analytical tasks and visual designs.
\system{} is constructed iteratively through a workshop study with 12 VA designers.
The structure provides opportunities to understand and utilize state-of-the-art visualization designs, which are demonstrated through a user study.
However, given that designing a visual analytics system is a complex procedure, 
we note that our work is the first step toward understanding and indexing VA systems.
We hope that our \system{} and lessons learned could provide a helpful foundation for further research.

\begin{acks}
The work was supported by NSF of China (U22A2032) and the Collaborative Innovation Center of Artificial Intelligence by MOE and Zhejiang Provincial Government.
The work was also partially supported by the Ministry of Education, Singapore, under its Academic Research Fund Tier 2 (Proposal ID: T2EP20222-0049).
\end{acks}

\bibliographystyle{ACM-Reference-Format}
\bibliography{main}

\end{document}